# Accelerated rare event sampling

David Yevick

Department of Physics

University of Waterloo

Waterloo, ON N2L 3G7

**Abstract:** A sampling procedure for the transition matrix Monte Carlo method is introduced that generates the density of states function over a wide parameter range with minimal coding effort.

**Introduction:** Standard Monte-Carlo simulations of physical systems simulate the collective behavior of physical systems quantified by one or more (typically macroscopic) system variables $\vec{E}(\vec{\alpha})$ by assigning random values to the underlying (microscopic) parameters, $\vec{\alpha}$. While such a procedure is straightforward, physically significant rare events are inefficiently generated. Accordingly, importance sampling [1], multicanonical [2] [3] [4] [5] [6] and Wang-Landau [7] [8] methods enhance the probability of these events either computationally or experimentally [9] through biased sampling. Briefly, in these procedures, specializing for simplicity to a single system variable, small, random changes in $\vec{\alpha}$ are accepted in accordance with a rule that favors displacements toward low probability regions of $E$. In the multicanonical method, this rule is updated iteratively by first generating an estimate, $p^{(1)}(E_i)$, of the density of states as a function of energy, $p(E)$, here normalized to unity - which coincides with the infinite temperature probability distribution function - through an unbiased Monte-Carlo calculation. Subsequently, an initial set of parameter values $\vec{\alpha}_{\text{current}}$, here assumed chosen from a limited and discrete set of possible values, is selected and then altered by a small, randomly generated quantity according to $\vec{\alpha}_{\text{new}} = \vec{\alpha}_{\text{current}} + \delta\vec{\alpha}$. The new realization is accepted with probability $\min\left(p^{(1)}(E_{\text{current}})/p^{(1)}(E_{\text{new}}),1\right)$ followed by a prescribed number of similar displacement and acceptance steps. The resulting histogram $n(E_i)$ of the number of realizations recorded in each interval $i$ of $E$ (where $n(E_i)$ is also incremented after a rejected move from interval $i$ to $j$) then exhibits the desired biased towards lower probability regions while preserving detailed balance. Multiplying $n(E_i)$ by $p^{(1)}(E_i)$, both of which are typically set to unity in unsampled histogram bins, yields a new probability distribution $p^{(2)}(E_i)$; the process is then iterated with $p^{(m)}(E)$ replaced by $p^{(m+1)}(E)$.

As the above procedure only samples low probability realizations after several iterations, Wang and Landau proposed retaining the multicanonical acceptance rule while multiplying the probability density in the *j*:th bin each time the bin is visited by a factor $\lambda^{(m)}$ that typically equals $ce^{1/m}$ for the *m*:th iteration of the method. While detailed balance is initially strongly violated, convergence is attained for large *m*.

The transition matrix formalism instead constructs a matrix $\mathbf{T}$ such that $T_{ij}$ corresponds to the *probability that a realization in a histogram bin $E_i$ transitions to bin $E_j$ after the displacement $\delta\vec{\alpha}$*. The normalized eigenvector of $\mathbf{T}$ with unit eigenvalue then coincides with the desired probability distribution $p(E)$. *The transition matrix records all accepted and rejected transitions, hence arbitrary acceptance rules that do not preserve detailed balance can be employed.* These include the multicanonical and Wang-Landau rule [10] [11] [12] [13] [14] as well as rules based on properties of the transition probability between microscopic states [15] [16] or the ratio of transition matrix elements [17] [18] [16]. Alternatively, [11] [19] [20] limited transitions from the *i*:th bin to bins that have been previously visited a smaller number of times, insuring that the number of realizations recorded in each bin is nearly independent of $i$.

**Numerical procedure:** This paper proposes a new strategy for transition matrix calculations that both most efficiently visits all desired $E$ values and estimates transition probabilities between states with equal relative accuracy in both low and high probability regions. As well, paths between differing low probability regions are sampled at regular intervals. In particular, starting with a randomly generated realization, $\alpha$, for which the value of $E$ falls in the $i$:th histogram bin *perturbed realizations are rejected until a transition occurs to a bin j with $j \geq i$*. Subsequently, only a transition to a bin $k \geq j$ is accepted and this procedure is continued until either the last of the $N$ bins comprising the computational window is sampled or a realization occurs for which none of the possible perturbations (when these are discrete and limited in magnitude) yield a state with larger bin index. The procedure is then repeated but in the direction of decreasing bin number. Examples of perturbations that are not limited in magnitude and can therefore always escape from local extrema are, for example, those that change each system variable with a certain probability or that act on a single variable but possess a magnitude described by an unbounded (e.g. Gaussian) distribution.

Once the transition matrix is constructed, the probability distribution is determined with either of two procedures. In the first a random vector, $x^{(0)}$, here constructed with elements uniformly distributed in the interval $[0,1]$, is multiplied repeatedly by the transition matrix, $\mathbf{T}$. Since the unit eigenvalue is the largest eigenvalue of this matrix the eigenvector corresponding to the state density is obtained after a sufficient number of multiplications. [19] Alternatively, the system of equations $(\mathbf{T}-\mathbf{I})x^{(m+1)} = x^{(m)}$ where $\mathbf{I}$ denotes the unit matrix can be repeatedly solved. The latter method is however sensitive to numerical error as $\mathbf{T}-\mathbf{I}$ is nearly singular.

**Computer program:** A fully functional modified transition matrix Octave/MATLAB program written with the syntax (e.g. spacing and naming) conventions of [21] [22] for the probability distribution, $p(E)$, for the number of heads, $E$ associated with a system of $N_{\text{coin}} = 400$ coins is given below. While such a non-interacting system is physically trivial, the code can be easily understood and adapted to a wide variety of problems. Here $2 \times 10^5$ realizations are employed together with 1000 and 6 iterations for the multiplicative and linear equation solvers, respectively.

```
clear all;
numberOfCoins = 400;
```

```
histogramR = ones( 1, numberOfCoins + 1 );
transitionMatrixRC = zeros( numberOfCoins + 1, numberOfCoins + 1 );
% Initial realization
realizationR = round ( rand( 1, numberOfCoins ) );
histogramBin = sum( realizationR ) + 1;
newRealizationR = realizationR;
numberOfRealizations = 200000;
directionFlag = 1;
for realizationInstance = 1 : numberOfRealizations
    % Coin flip
    coinNumber = ceil( numberOfCoins * rand( ) );
    newRealizationR(coinNumber) = 1 - realizationR(coinNumber);
    newHistogramBin = sum( newRealizationR ) + 1;
    transitionMatrixRC(histogramBin, newHistogramBin) = transitionMatrixRC(histogramBin, newHistogramBin) + 1;
    % Acceptance rule
    if ( newHistogramBin > numberOfCoins | newHistogramBin < 2 ) directionFlag = -directionFlag; end;
    if ( ( directionFlag == 1 & newHistogramBin >= histogramBin ) | ...
        ( directionFlag == -1 & newHistogramBin <= histogramBin ) )
        realizationR(coinNumber) = newRealizationR(coinNumber);
        histogramBin = newHistogramBin;
    else
        newRealizationR(coinNumber) = realizationR(coinNumber);
    end
    histogramR(histogramBin) = histogramR(histogramBin) + 1;
end
for rowIndex = 1 : numberOfCoins
    sumRow = max( sum( transitionMatrixRC(rowIndex, :) ), 1);
    transitionMatrixRC(rowIndex, :) = transitionMatrixRC(rowIndex, :) / sumRow;
end
methodChoice = 1;
randomVectorR = rand( 1, numberOfCoins + 1);
% Matrix multiplication
if ( methodChoice == 1 )
    numberOfMultiplications = 1000;
    for multiplicationLoop = 1 : numberOfMultiplications
        randomVectorR = randomVectorR * transitionMatrixRC;
        randomVectorR = randomVectorR / sum( randomVectorR );
    end
else
% Iterative equation solver
    numberOfIterations = 6;
    modifiedTransitionMatrixRC = transitionMatrixRC - eye(numberOfCoins + 1, numberOfCoins + 1;
    for iterationLoop = 1 : numberOfIterations
        randomVectorR = randomVectorR / modifiedTransitionMatrixRC;
        randomVectorR = randomVectorR / sum( randomVectorR );
    end
```

    **end**

    To determine efficiently the state density of e.g. the two-dimensional Ising model, the above code should be somewhat modified. In particular, the procedure for generating each new state should ensure that the calculation can always exit from the statistically rare states for which reversing any single spin yields a realization with higher or lower energy. Method A changes the spin of a site with a random probability in such a manner that on average only a number **averageNumberOfSpInFlips** (below 1 and 4 for the 5x5 and 40x40 Ising model respectively) of spins are altered:

**for spinNumber = 1 : numberOfSpinsSquared;**
    **if ( rand < averageNumberOfSpinFlips / numberOfSpinsSquared )**
        **newRealizationR(spinNumber) = -realizationR(spinNumber);**
    **end**
**end**

Alternatively, in method B starting from a given state, new realizations can be generated by inverting each spin separately in a constantly changing random order. If none of these realizations lead to a state with a change in $E$ in the desired direction, the direction of acceptance in $E$ is reversed. The relevant code lines for the 40x40 Ising model in which only configurations with more than 500 upward pointing spins and fewer than 500 downward pointing spins are considered can be written

**loopCount = mod( loopCount, numberOfSpinsSquared ) + 1;**
**if ( loopCount == 1 )**
    **selectionVector = 1 : numberOfSpinsSquared;**
    **for loop = numberOfSpinsSquared : -1 : 2**
        **vectorIndex = randi( loop - 1 );**
        **tempResult = selectionVector( loop );**
        **selectionVector( loop ) = selectionVector( vectorIndex );**
        **selectionVector( vectorIndex ) = tempResult;**
    **end**
**end;**
**spinNumber = selectionVector( loopCount );**
**newRealizationR(spinNumber) = - realizationR(spinNumber);**
**newHistogramBin = costFunctionIsing2d( newRealizationR );**
**transitionMatrixRC(histogramBin, newHistogramBin) = transitionMatrixRC(histogramBin, newHistogramBin) + 1;**
**if ( ( directionFlag == 1 & newHistogramBin >= histogramBin ) | ...**
                **( directionFlag == -1 & newHistogramBin <= histogramBin ) )**
**if ( newHistogramBin >= numberOfSpins2 - 500 | newHistogramBin <= 500 + 1 ) directionFlag = - directionFlag; end;**
    **realizationR = newRealizationR;**
    **histogramBin = newHistogramBin;**
    **loopFlag = 1;**
**else**
    **if ( loopFlag == numberOfSpinsSquared ) directionFlag = - directionFlag; loopFlag = 1; end;**
    **newRealizationR = realizationR;**
    **loopFlag = loopFlag + 1;**

end

**Numerical results:** To compare the above method to previous techniques, we first display the result of a standard Wang-Landau *transition matrix* calculation of the probability distribution function associated with 200 random coin flips associated with 200,000 realizations containing all accepted and rejected transitions resulting from a single iteration with a multiplication factor $\lambda = e^{1/2}$. Identifying $E$ with the number of heads, and applying the matrix multiplication procedure to determine $p(E)$, the results for the standard unbiased Monte-Carlo method, (dashed line), a single iteration of the Wang-Landau approach (dotted line) and the combined single iteration Wang-Landau transition matrix simulation (dashed-dotted line marked +) are superimposed on the exact values (solid line) in Figure 1. As expected, incorporating the transition matrix formalism greatly increases the accuracy of the Wang-Landau calculation. The program of the previous section in contrast yields the results of Figure 2 and Figure 3 for the matrix multiplication and iterative equation solution procedures respectively. In both cases, exact and numerical results agree for all $E$. The associated number of realizations registered in each bin, displayed in Figure 4, reflects the factor of 200 greater probability of a directed transition from a $E = 200$ realization compared to the $E = 1,399$ states.

The two dimensional Ising model with zero external magnetic field, periodic boundary conditions, and an unit amplitude antiferromagnetic interaction between half-integral spins affords a non-trivial example of the above procedure. The energy variable, $E$, in the calculation is rescaled so that the lowest possible value is zero while the possible states are separated by unity by dividing the sum of the energy and the square of the number of spins by two. The normalized density of states, $p(E)$, for a 5x5 spin lattice calculated with method A with **numberOfRealizations** = 400,000 realizations is compared in Figure 5 to the exact result (solid line) and the first and second 500,000 sample iterations of the multicanonical algorithm (downward and upward pointing triangles) with an identical acceptance rule. Since the results effectively coincide, the identical multicanonical and transition calculations for a 40x40 spin lattice are subsequently compared. Here method A is again employed while the direction of the transitions is reversed whenever $E$ is displaced by 500 from one of the two boundaries of the computational window. The two computational procedures are again in excellent agreement, c.f. Figure 6; however, the transition matrix method estimates the density of states over a wide region of $E$ with relatively few realizations. A graph of the number of recorded realizations for each value of $E$ in the transition matrix method is finally presented in Figure 7, which exhibits the expected increase in sampling frequency for less probable system parameter values.

**Discussion and conclusions:** The relative advantages and disadvantages of the proposed method are most clearly evident in the context of the coin flip example. Consider for example transitions from a state with $E = 1$. Since the transition to a $E = 2$ state is 399 times more likely than the $E = 0$ transition, the system will on average remain in a $E = 1$ state or states for 399 realization steps before transitioning to $E = 0$. While during these steps the elements $T_{1j}$ for only $E = 1$ realizations are accumulated, any method that permits transitions to $E = 2$ requires on average considerably more steps before the $E = 0$ state is sampled (after an $E = 2$ transition returning to $E = 1$ involves at least an average of 200 additional steps for any acceptance rule). However $\approx 200$ steps are necessary to determine the transition probability from a given $E = 1$ to a $E = 0$ state to the same level of relative accuracy as the probability from a $E = 200$ to a $E = 201$ state. The procedure above is consequently optimal if the physically significant realizations possess very low probability. The transition rule can as well be altered during the calculation to favor physically interesting or undersampled regions of $E$ identified from the intermediate realizations or results. For example, if an interval $a < E < b$ such as e.g. the region around the outage boundary in communication system calculations [11] is determined to be particularly significant, the range over which the realizations are sampled can be temporarily or permanently limited to this interval.

While the above transition matrix procedure is thus more flexible, easily implemented and potentially faster than previously proposed techniques, its relative accuracy is problem specific. Further, geometry-related issues arise in the calculation of multidimensional density of states functions of several macroscopic parameters $\vec{E}$. For example, in two dimensions, the procedure of this paper can be applied twice to first obtain realizations within each bin of $E_1$ at a fixed $E_2$ after which these of realizations can be employed to sample each $E_1$ for all other values of $E_2$. However, such strategies entail a greater percentage of rejected transitions, far larger transition matrices and more involved code than in one dimension.

**Acknowledgements:** The Natural Sciences and Engineering Research Council of Canada (NSERC) is acknowledged for financial support.

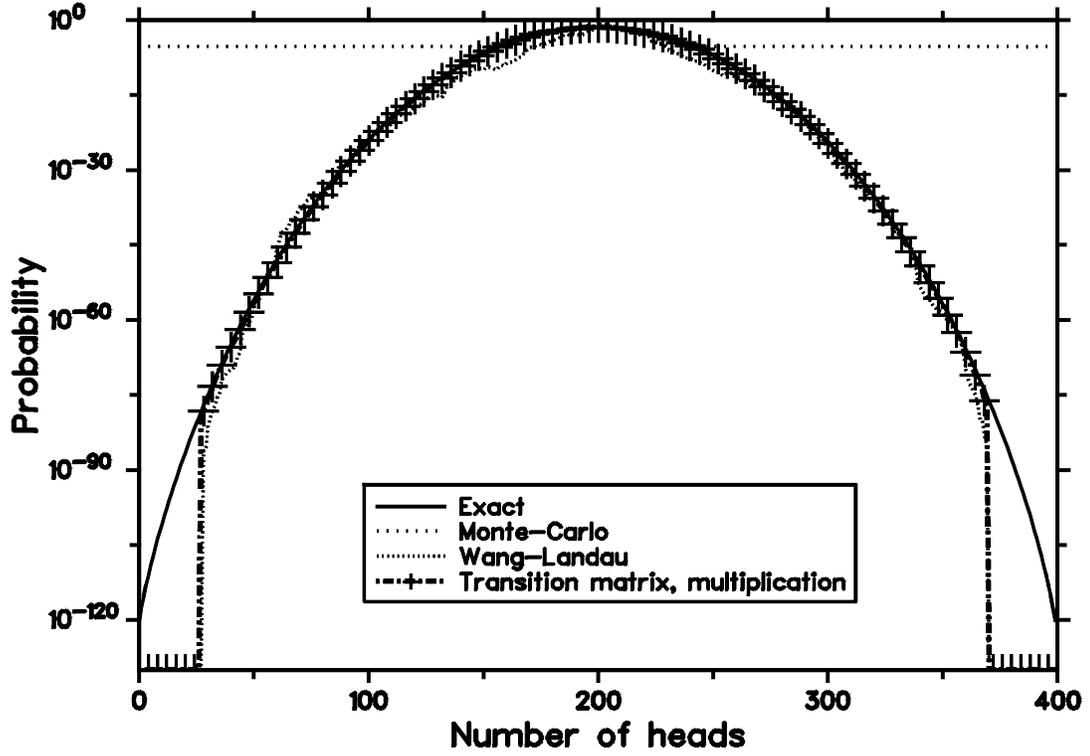

Figure 1: The exact, Monte-Carlo, Wang-Landau and Wang-Landau transition matrix results for the probability of $N$ heads after 400 coin flips

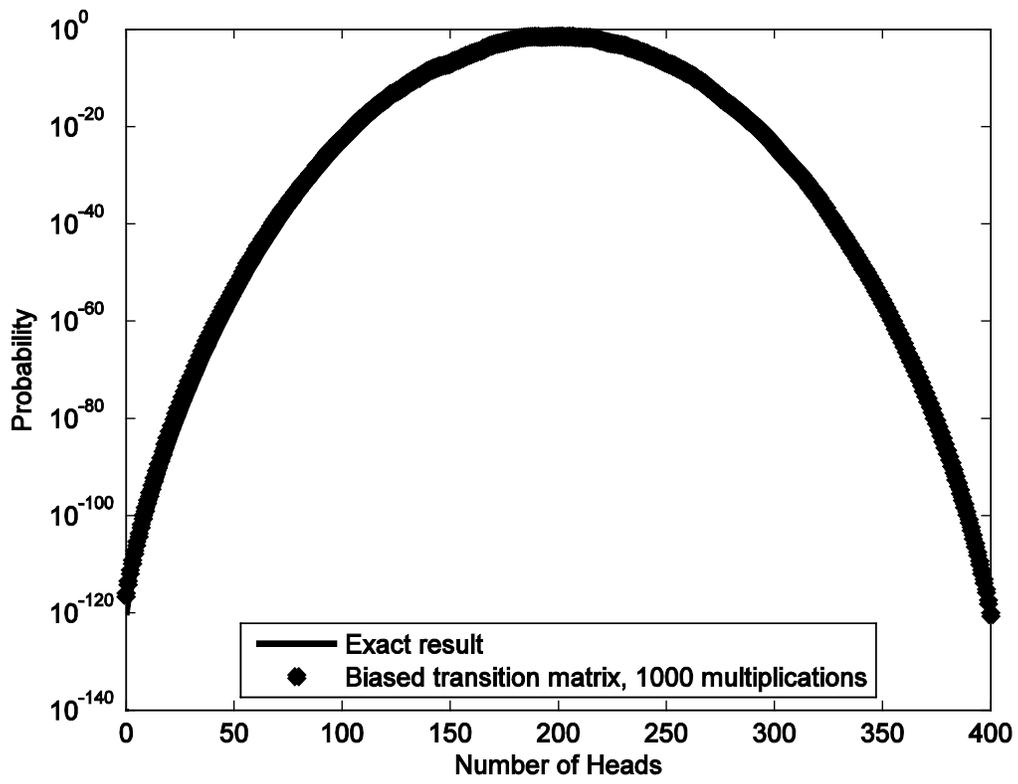

*Figure 2: The exact and simplfied transition matrix results generated with the multiplicative solver for the problem of Figure 1*

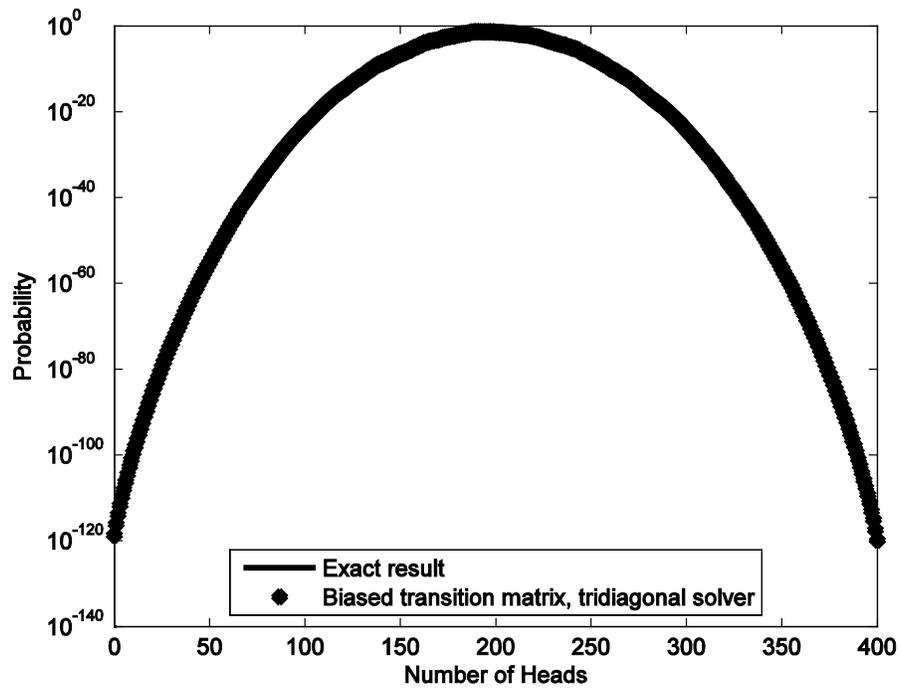

*Figure 3: The exact and simplified transition matrix results generated with the linear equation solver for the problem of Figure 1*

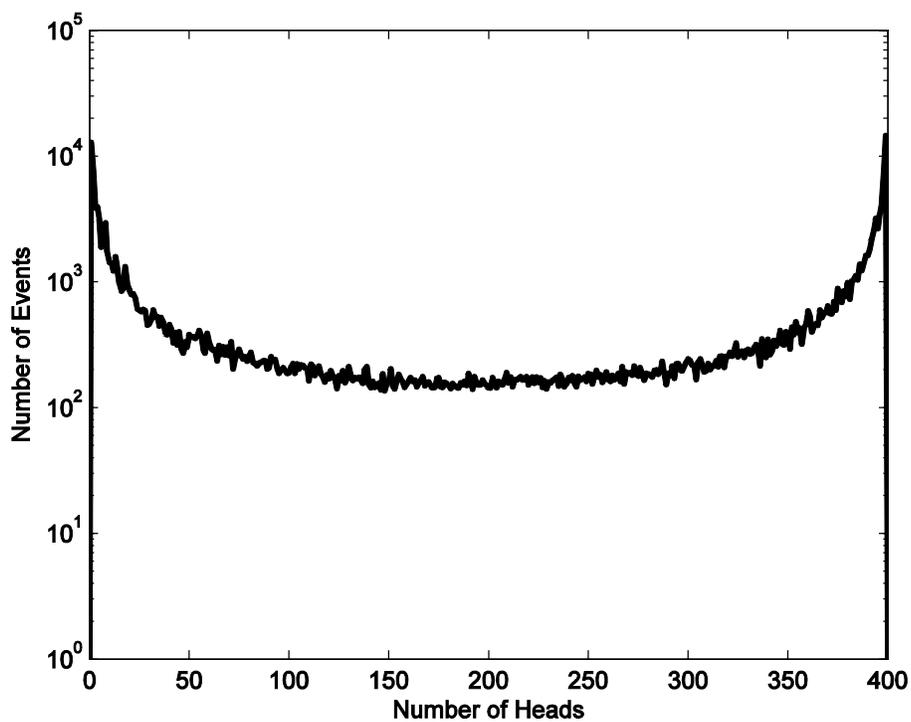

Figure 4: Number of realizations in each histogram bin for the transition matrix result of the preceding figures

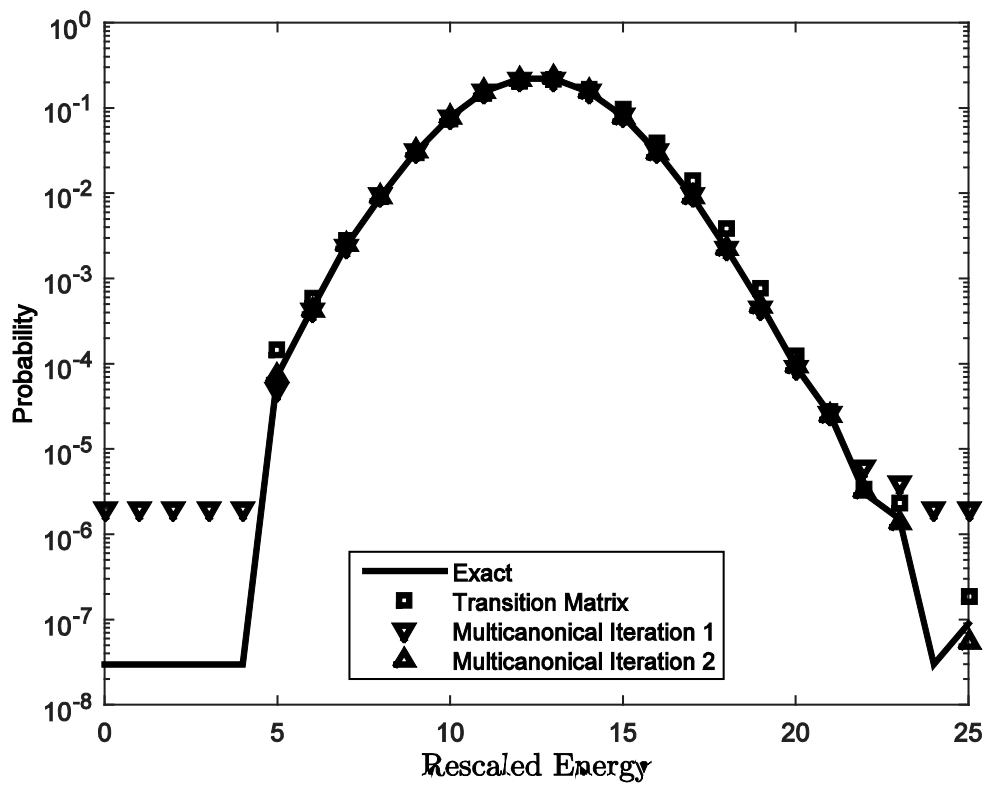

*Figure 5: The normalized density of states for the two-dimensional 5x5 ising model as calculated with the exact, transition matrix and multicanonical methods*

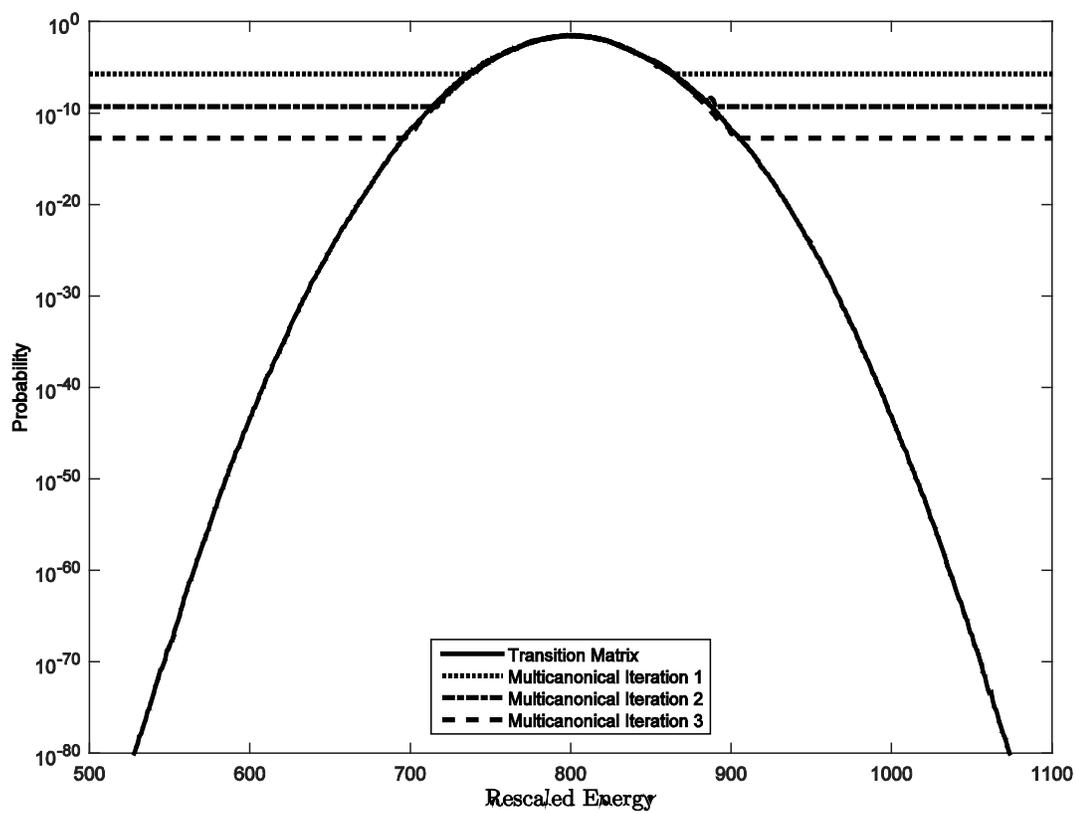

*Figure 6: The transition matrix density of states (solid line) and the results for the first three iterations of the multicanonical method for the 40x40 ising model*

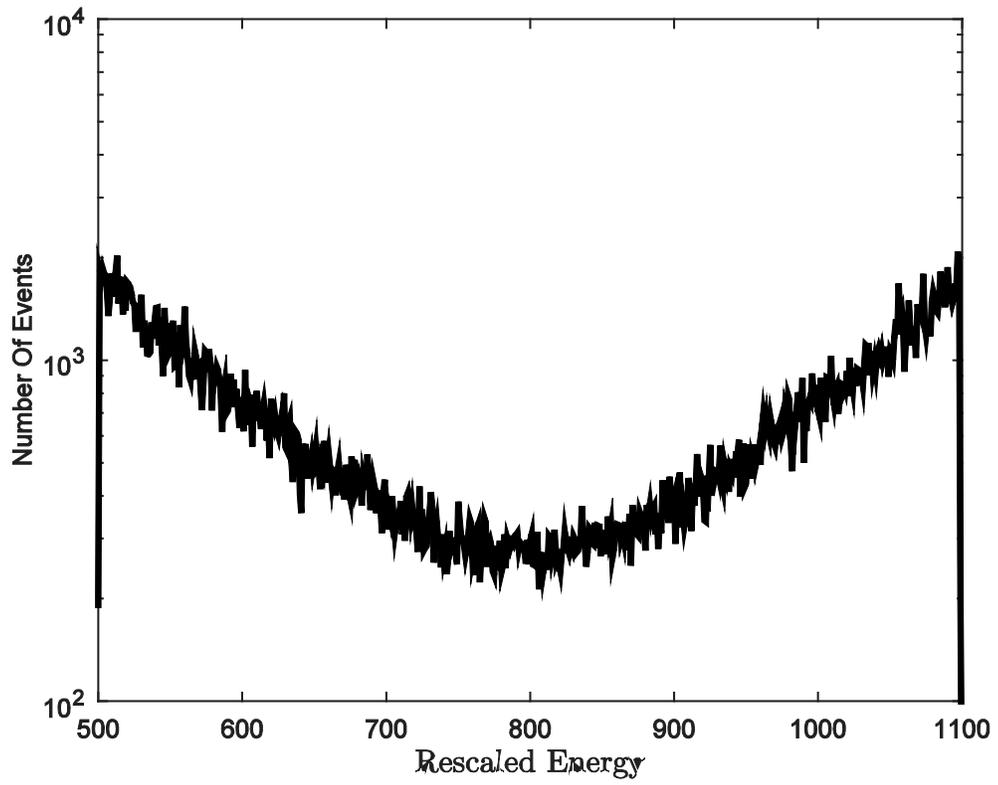

Figure 7: The number of occurrences of state in the transition matrix of the preceeding figure